\documentclass[a4paper,12pt]{article}

\usepackage{amsmath,amssymb,amsfonts}
\usepackage{graphicx,epsfig, subfigure}
\usepackage{hyperref}
\usepackage{color}
\usepackage{amsmath}
\usepackage{amsfonts}
\usepackage{amssymb}
\usepackage{graphicx}
\usepackage{slashed}
\usepackage{epigraph}
\numberwithin{equation}{section}
\definecolor{red}{cmyk}{0,1,1,0.4}

\title{Experimental Constraints on Heavy Fermions\\ in Higgsless Models}
\date{}
\author{Raffaele Tito D'Agnolo}

\begin{document}
\maketitle
\begin{center}
 \textit{Scuola Normale Superiore, Piazza dei Cavalieri 7, 56126 Pisa, Italy}
\end{center}
\mbox{} \newline
\begin{abstract}
Using an effective Lagrangian approach we analyze a generic 
Higgsless model with composite heavy fermions, 
transforming as $SU(2)_{L+R}$ Doublets. Assuming that 
the Standard Model fermions acquire mass through mixing with the 
new heavy fermions, we  constrain the free parameters of the 
effective Lagrangian studying Flavour Changing Neutral Current 
processes.
In so doing we obtain bounds that can be applied to a wide range 
of models characterized by the same fermion mixing hypothesis.
\end{abstract}
\mbox{}
\newpage
\section{Introduction}
The final word on the mechanism of electroweak symmetry breaking (EWSB) has not yet been written, since we have been lacking direct experiments at the relevant energy scale. This is one of the reasons why the models, that have been presented so far, do not appear fully satisfying. They are usually confronted with conceptual difficulties and often need \textit{ad hoc} solutions or some amount of fine tuning to account for the ElectroWeak Precision Tests (EWPT) and the measurements in flavour physics. In addition to that not a single candidate has characteristic features that make it clearly favoured with respect to the others. The only possible exception is the Higgs mechanism~\cite{Higgs:1966ev, Englert:1964et,Kibble:1967sv, Guralnik:1964eu}, that, however, has to cope with the hierarchy problem and the little hierarchy problem, not to mention the tension between the light Higgs boson favoured by the EWPT~\cite{:2004qh} and the unsuccessful direct searches performed so far.

This brief account of present difficulties in finding a solid mechanism of electroweak symmetry breaking encourages a rather model-independent analysis. Furthermore in the absence of direct evidence of the existence of an Higgs boson, it is still legitimate to explore \textit{Higgsless} approaches to the 
problem (see e.g.~Ref.~\cite{Csaki:2003dt,Nomura:2003du,Barbieri:2003pr,Foadi:2003xa,Georgi:2004iy,Chivukula:2004pk,Barbieri:2008cc}).

With this spirit we consider a model based on an electroweak chiral 
Lagrangian. In particular, 
following the proposal of Ref.~\cite{Barbieri:2008zt}, we focus 
on the consequences of giving masses to the Standard Model fermions 
through mixing with Composite Fermions.
This is a strong assumption of the model and the aim of this work is precisely to test it using the data acquired in the past years in flavour physics.

The main features of the effective Lagrangian framework we are considering 
are similar to those discussed in Ref.~\cite{Barbieri:2008cc,Barbieri:2008zt}. 
We assume that some strong dynamics breaks  a 
$SU(2)_{L}\times SU(2)_{R}\times U(1)_{X}$ symmetry down to $SU(2)_{L+R}\times U(1)_{X}$. The symmetry group is global in the limit of 
vanishing electroweak gauge couplings and the hypercharge 
is given by $Y=T_{3R}+X$. This spontaneous symmetry breaking, 
that occurs at the Fermi scale $v=(\sqrt{2}G_{F})^{-1/2}\approx 246$~GeV, 
leads also to the breaking of the electroweak symmetry 
$SU(2)_{L}\times U(1)_{Y}\to U(1)_{e.m.}$.

It is assumed that the strong dynamics generates composite vectors and composite fermions with definite transformation properties under the coset group $SU(2)_{L}\times SU(2)_{R}/SU(2)_{L+R}$. We parametrize it through the Goldstone fields $\hat{\pi}(x)=\pi^{a}(x)\sigma^{a}/2$, according to the CCWZ formalism~\cite{Coleman:1969sm,Callan:1969sn}.
The role of the composite vectors was described in detail in~\cite{Barbieri:2008cc} and amounts to keep unitarity in the elastic and inelastic scattering amplitudes of the gauge bosons up to a scale $\Lambda \approx 4\pi v\approx 3$~TeV.

The real focus of this work are the composite fermions. 
In Ref.~\cite{Barbieri:2008zt} the case in which the composite fermions are $SU(2)_{L+R}$ singlets was described in detail, here we 
assume them to be $SU(2)_{L+R}$ doublets. We also assume they appear 
in three generations and with the appropriate $X$ quantum number 
that allows the mixing with Standard Model fermions (that henceforth we 
call \textit{elementary}). As mentioned above, a key assumption is that, 
before mixing occurs, the elementary fermions are massless.
What makes the Doublets interesting is that, any reasonable breaking pattern of the flavour symmetry arising from the mixing, does not reproduce the CKM picture of flavour physics. The same is not true for Singlets~\cite{Barbieri:2008zt}, where the CKM description can be recovered. The emergence of this new flavour structure allows to set stringent constraints on the free parameters of the model.

The paper is organized as follows. After briefly presenting the 
Composite Fermions model in section \ref{s2}, we turn to the computation 
of FCNC effective Lagrangians in section \ref{s3}. The constrain 
on the the free parameters of the model are analyzed in section \ref{s4}.

\section{The Composite Fermions model}\label{s2}
In this section, for the sake of completeness, we 
briefly review the main characteristics of the model already presented in \cite{Barbieri:2008zt}.
\subsection{The effective Lagrangian}
It is convenient to introduce first the Goldstone Fields
\begin{equation}
 U\equiv e^{i2\hat{\pi}/v}, \quad \hat{\pi}=\pi^{a}T^{a}=\pi^{a}\sigma^{a}/2=\left(\begin{array}{cc}
                                                                 \frac{\pi^{0}}{\sqrt{2}}&\pi^{+}\\
                                                                  \pi^{-}&-\frac{\pi^{0}}{\sqrt{2}}
                                                                \end{array}\right),
\end{equation}
 transforming under $SU(2)_{L}\times SU(2)_{R}$ as
\begin{equation}
 U'(x)=g_{L}U(x)g_{R}^{\dagger}.
\end{equation}
We also define 
\begin{equation}
 u^{2}\equiv U, \qquad  u'=g_{R}uh^{\dagger}=hug_{L}^{\dagger}, \qquad h\in SU(2)_{L+R}
\end{equation}
and the functions
\begin{equation}
 \Gamma_{\mu}=\frac{1}{2}\left[u^{\dagger}(\partial_{\mu}-i\hat{B}_{\mu})u+u(\partial_{\mu}-i\hat{W}_{\mu})u^{\dagger}\right], \quad \Gamma_{\mu}^{\dagger}=-\Gamma_{\mu}, \quad \Gamma_{\mu}'=h\Gamma_{\mu}h^{\dagger}+h\partial_{\mu}h^{\dagger}
\end{equation}
\begin{equation}
 u_{\mu}=iu^{\dagger}D_{\mu}Uu^{\dagger}=u_{\mu}^{\dagger}, \quad u_{\mu}'=hu_{\mu}h^{\dagger},
\end{equation}
At this point we are ready to write down the effective Lagrangian of 
the composite fermions 
\begin{displaymath}
 \mathcal{D}\equiv\left(\begin{array}{c}
                T \\ B
               \end{array}\right),
\end{displaymath}
which have $X$-number $1/6$
and transform under $SU(2)_{L+R}$ as $\mathcal{D}'=h\mathcal{D}$. 
Assuming that the strong dynamics conserves parity,
the most general lowest-order Lagrangian bilinear 
in the $\mathcal{D}$ field is 
\begin{equation}
 L_{\mathcal{D}}=i\overline{\mathcal{D}}\gamma^{\mu}\left(\partial_{\mu}+\Gamma_{\mu}-ig'XB_{\mu}\right)\mathcal{D}+\frac{\alpha}{2}\overline{\mathcal{D}}\gamma^{\mu}\gamma_{5}u_{\mu}\mathcal{D}+M_{\mathcal{D}}\overline{\mathcal{D}}\mathcal{D},
\label{eq:LD}
\end{equation}
where $\alpha$ is a free parameter of order 1.
The gauge invariant Lagrangian describing the mixing with the elementary fields reads
\begin{equation}
 L^{\mathcal{D}}_{mix}=m^{u}_{L}\overline{\mathcal{D}}_{R}u^{\dagger}\hat{P}_{u}Uq_{L}+m^{u}_{R}\overline{\mathcal{D}}_{L}u^{\dagger}\hat{P}_{u}q_{R}+m^{d}_{L}\overline{\mathcal{D}}_{R}u^{\dagger}\hat{P}_{d}Uq_{L}+m^{d}_{R}\overline{\mathcal{D}}_{L}u^{\dagger}\hat{P}_{d}q_{R}+h.c. \label{flavDou}
\end{equation}
and it is responsible for the breaking of the flavour symmetry.

\subsection{The flavour symmetry}
We assume, again following~\cite{Barbieri:2008zt}, that in absence of mixing the model possesses a large flavour symmetry which includes that of the SM in the limit of vanishing Yukawa couplings,
\begin{equation}
 G_{f}^{SM}=SU(3)_{q}\times SU(3)_{uR}\times SU(3)_{dR},
\end{equation}
together with the flavour symmetry of the composite sector. So we have
\begin{equation}
 G_{f}^{\mathcal{D}}=SU(3)_{\mathcal{D}}\times SU(3)_{q}\times SU(3)_{uR}\times SU(3)_{dR}, \label{gfd}
\end{equation}
which must be broken in order to be consistent with experiments. In the SM, viewed as an effective theory, $G_{f}^{SM}$ is broken by two dimensionless parameters $Y^{u}$ and $Y^{d}$, which treated as spurions, transform as
\begin{equation}
 Y^{u}=(3,\overline{3}) \quad \mathrm{under} \quad SU(3)_{uR}\times SU(3)_{q},
\end{equation}
\begin{equation}
 Y^{d}=(3,\overline{3}) \quad \mathrm{under} \quad SU(3)_{dR}\times SU(3)_{q}.
\end{equation}
This hypothesis \cite{Chivukula:1987py,Cirigliano:2005ck,D'Ambrosio:2002ex} enforces the CKM picture: without loss of generality $Y^{d}$ can be reduced to diagonal form and $Y^{u}$ can be diagonalized up to a single unitary matrix. Even the inclusion of higher dimensional operators, suppressed by a scale of $3$ to $5$ TeV, does not undermine the consistency with experimental data, provided that this symmetry breaking pattern is respected~\cite{D'Ambrosio:2002ex}.

There are no suitable symmetry conditions that can restore the CKM picture 
in the case of Doublets. As discussed in~\cite{Barbieri:2008zt}, the two 
most promising possibilities are the so called Parity Conserving (PC) 
and Parity Breaking (PB) patterns. In the Parity Conserving  case 
the Yukawa parameters transform as
\begin{equation}
 Y^{u}_{3}=(3,\overline{3}) \quad \mathrm{under} \quad SU(3)_{\mathcal{D},\mathcal{T}}\times SU(3)_{q+uR},
\end{equation}
\begin{equation}
 Y^{d}_{3}=(3,\overline{3}) \quad \mathrm{under} \quad SU(3)_{\mathcal{D},\mathcal{T}}\times SU(3)_{q+dR}
\end{equation}
and the mixing Lagrangian can be written as
\begin{equation}
 L_{mix}^{\mathcal{D}}(PC)=v\overline{U}_{R}\mathcal{V}\lambda^{u}Vu_{L}+f^{u}\overline{U}_{L}\mathcal{V}\lambda^{u}u_{R}+v\overline{D}_{R}\lambda^{d}d_{L}+f^{d}\overline{D}_{L}\lambda^{d}d_{R}+ h.c. \;.
\end{equation}
Here  $V$ is the usual CKM matrix, while $\mathcal{V}$ is a further unitary matrix that can not be eliminated through symmetry considerations.\footnote{ 
Note that there is a typo in the expression of $L_{mix}^{\mathcal{D}}(PC)$ 
reported in \cite{Barbieri:2008zt}.}
In the  Parity Breaking case $G_{f}^{\mathcal{D}}$ is broken down to $SU(3)_{\mathcal{D}+uR}~\times SU(3)_{\mathcal{D}+dR}\times SU(3)_{q}$, which in turn is only broken by
\begin{equation}
 Y^{u}_{2}=(3,\overline{3}) \quad \mathrm{under} \quad SU(3)_{\mathcal{D}+uR}\times SU(3)_{q},
\end{equation}
\begin{equation}
 Y^{d}_{2}=(3,\overline{3}) \quad \mathrm{under} \quad SU(3)_{\mathcal{D}+dR}\times SU(3)_{q}.
\end{equation}
Therefore the mixing Lagrangian reads  
\begin{equation}
 L_{mix}^{\mathcal{D}}(PB)=v\overline{U}_{R}\mathcal{V}\lambda^{u}Vu_{L}+f^{u}\overline{U}_{L}u_{R}+v\overline{D}_{R}\lambda^{d}d_{L}+f^{d}\overline{D}_{L}d_{R}+ h.c. \;.
\end{equation}
\subsection{The fermion mass eigenstates}\label{s23}


It is easy to verify, that after sending $u_{L}\rightarrow V^{\dagger}u_{L}$ and $U[U_{R}] \rightarrow \mathcal{V}U[U_{R}]$ (in the PC[PB] case), the mass matrix for both, up- and down-type quarks, reduces to three $2\times 2$ blocks, labeled by a generation index,
\begin{displaymath}
 M^{(u)}_{i}=\left(\begin{array}{cc}
              0 & f^{u}\lambda^{u}_{i}[f^{u}]\\
  v\lambda^{u}_{i} & M_{U_{i}}
             \end{array}\right) \quad 
M^{(d)}_{i}=\left(\begin{array}{cc}
              0 & f^{d}\lambda^{d}_{i}[f^{d}]\\
  v\lambda^{d}_{i} & M_{D_{i}}
             \end{array}\right), 
\end{displaymath}
where, again, the values outside/within square brackets correspond to the PC/PB case of flavour symmetry breaking. Note that $M_{D_{i}}=M_{U_{i}}=M_{\mathcal{D}_{i}}$.

 This introduces two extra parameters for each quark with respect to the SM, that can be chosen as the mass of the heavy partner and the mixing angle in the left-handed sector.\\
The mixing angles are defined as follows,
\begin{equation}
  \left( \begin{array}{cc}
-c_{R}^{q} & s_{R}^{q} \\
s_{R}^{q} & c_{R}^{q} \\
\end{array} \right)
 \left( \begin{array}{cc}
0 & m_{R}^{q} \\
m_{L}^{q} & M_{\mathcal{D}_{q}} \\
\end{array} \right)
 \left( \begin{array}{cc}
c_{L}^{q} & s_{L}^{q} \\
-s_{L}^{q} & c_{L}^{q} \\
\end{array} \right)=
 \left( \begin{array}{cc}
m_{q} & 0 \\
0 & M_{Q} \\
\end{array} \right), \label{mmatrix}
\end{equation}
where we have considered a generic $2\times 2$ block. It is worth noticing that in the limit of vanishing light quark masses, the left-handed mixing angles are zero and the heavy degrees of freedom decouple in low energy observables. From the previous definition we can obtain the very useful relations
\begin{equation}
 t_{R}^{q}t^{q}_{L}=\frac{m_{q}}{M_{Q}}, \quad m_{q}=\frac{m_{L}^{q}m_{R}^{q}}{M_{Q}} \label{tan}.
\end{equation}
From these equations\footnote{ 
Note that in \cite{Barbieri:2008zt} $m_{q}=\frac{m_{L}^{q}m_{R}^{q}}{M_{Q}}$ is reported as an approximate relation, valid only at first order in $\frac{m_{L}^{q}}{ M_{\mathcal{D}_{q}}}$, but it is actually exact as can be verified taking the determinant of equation (\ref{mmatrix}).} it is apparent that the only relevant effect of the mixing arises in the top sector, as can be seen also from the approximate expressions
\begin{equation}
 (s_{L}^{q})^{2}_{PC}\approx\frac{vm_{q}M_{\mathcal{D}_{q}}^{2}}{f^{q}M^{3}_{Q}}, \quad (s_{L}^{q})^{2}_{PB}\approx\frac{m^{2}_{q}M_{\mathcal{D}_{q}}^{2}}{(f^{q})^{2}M^{2}_{Q}}, \label{topmixing}
\end{equation}
obtained in the limit $m_{L}^{q}\ll M_{\mathcal{D}_{q}}$.
\subsection{Couplings to Gauge vectors and Goldstone bosons}\label{s24}
In addition to the SM-like current, the Doublets' Lagrangian contains an interaction term that must be taken into account. Therefore the currents coupled to W and Z can be written as the sum of two terms, one of which is just the SM current rotated to the new mass eigenstates.
\begin{eqnarray}
  (J^{\mu}_{W})^{PC}_{\mathcal{D}}&=&(J^{\mu}_{W})_{SM}+(J^{\mu}_{W})^{PC}_{2} \\ \nonumber
  (J^{\mu}_{W})^{PB}_{\mathcal{D}}&=&(J^{\mu}_{W})_{SM}+(J^{\mu}_{W})^{PB}_{2} \\ \nonumber
 (J^{\mu}_{Z})_{\mathcal{D}}&=&(J^{\mu}_{Z})_{S}+(J^{\mu}_{Z})_{2},
\end{eqnarray}
The SM contribution to the currents is the same both in the PC and PB cases even if we include in $(J^{\mu}_{V})_{SM}$ also the current associated to the $X$ quantum number. For the W, this mixed current, containing both the SM and $X$ interactions, reads
\begin{eqnarray}
 (J^{\mu}_{W})_{SM}&=&\frac{g}{\sqrt{2}}\sum_{i,j=1,3}V_{ij}[c^{u_{i}}_{L}c^{d_{j}}_{L}\overline{u}^{i}_{L}\gamma^{\mu}d^{j}_{L}+c^{u_{i}}_{L}s^{d_{j}}_{L}\overline{u}^{i}_{L}\gamma^{\mu}D^{j}_{L}+ \label{six}\\ \nonumber
 &+&s^{u_{i}}_{L}c^{d_{j}}_{L}\overline{U}^{i}_{L}\gamma^{\mu}d^{j}_{L}+s^{u_{i}}_{L}s^{d_{j}}_{L}\overline{U}^{i}_{L}\gamma^{\mu}D^{j}_{L}]
\end{eqnarray}
and for the Z
\begin{eqnarray}
 (J^{\mu}_{Z})_{SM}&=&\frac{g}{c_{W}}\sum_{i=1,6}\left[\left(T_{3}\right)_{i}\left((c^{i}_{L})^{2}\overline{q}^{i}_{L}\gamma^{\mu}q^{i}_{L}+(s^{i}_{L})^{2}\overline{Q}^{i}_{L}\gamma^{\mu}Q^{i}_{L}\right)\right]+\label{jzsm}
\\ \nonumber
 &+&\frac{g}{c_{W}}\sum_{i=1,6}\left[\left(T_{3}\right)_{i}\left(c^{i}_{L}s^{i}_{L}\overline{q}^{i}_{L}\gamma^{\mu}Q^{i}_{L}+c^{i}_{L}s^{i}_{L}\overline{Q}^{i}_{L}\gamma^{\mu}q^{i}_{L}\right)\right]-\\ \nonumber
 &-&\frac{g}{c_{W}}\sum_{i=1,6}Q_{i}s^{2}_{W}\left[\overline{q}^{i}\gamma^{\mu}{q}^{i}+\overline{Q}^{i}\gamma^{\mu}{Q}^{i}\right].
\end{eqnarray}
The form of the interactions coming from the Doublets' Lagrangian, on the contrary, depends on the flavour symmetry breaking pattern. In fact, in the PC case we have
\begin{eqnarray}
 (J^{\mu}_{W})^{PC}_{2}&=&\frac{g(1+\alpha)}{2\sqrt{2}}\left\{\sum_{i,j=1,3}\mathcal{V}^{*}_{ij}\left[c^{u_{j}}_{L}\overline{U}^{j}_{L}-s^{u_{j}}_{L}\overline{u}^{j}_{L}\right]\gamma^{\mu}\left[-s^{d_{i}}_{L}d^{i}_{L}+c^{d_{i}}_{L}D^{i}_{L}\right]\right\}+ \label{eight}
\\ \nonumber
 &+&\frac{g(1-\alpha)}{2\sqrt{2}}\left\{\sum_{ij=1,3}\mathcal{V}_{ji}^{*}\left[c^{u_{j}}_{R}\overline{U}^{j}_{R}+s^{u_{j}}_{R}\overline{u}^{j}_{R}\right]\gamma^{\mu}\left[s^{d_{i}}_{R}d^{i}_{R}+c^{d_{i}}_{R}D^{i}_{R}\right]\right\},
\end{eqnarray}
while, in the PB case,
\begin{eqnarray}
 (J^{\mu}_{W})^{PB}_{2}&=&\frac{g(1+\alpha)}{2\sqrt{2}}\left\{\sum_{i=1,3}\left[c^{u_{i}}_{L}\overline{U}^{i}_{L}-s^{u_{i}}_{L}\overline{u}^{i}_{L}\right]\gamma^{\mu}\left[-s^{d_{i}}_{L}d^{i}_{L}+c^{d_{i}}_{L}D^{i}_{L}\right]\right\}+ \label{nine}
\\ \nonumber
 &+&\frac{g(1-\alpha)}{2\sqrt{2}}\left\{\sum_{ij=1,3}\mathcal{V}_{ji}^{*}\left[c^{u_{j}}_{R}\overline{U}^{j}_{R}+s^{u_{j}}_{R}\overline{u}^{j}_{R}\right]\gamma^{\mu}\left[s^{d_{i}}_{R}d^{i}_{R}+c^{d_{i}}_{R}D^{i}_{R}\right]\right\}
\end{eqnarray}
and
\begin{eqnarray}
 (J^{\mu}_{Z})_{2}&=&\frac{g}{2c_{W}}\sum_{i=1,6}(T^{3})_{i} \left[c^{i}_{L}\overline{Q}^{i}_{L}-s^{i}_{L}\overline{q}^{i}_{L}\right]\gamma^{\mu}\left[-s^{i}_{L}q^{i}_{L}+c^{i}_{L}Q^{i}_{L}\right]+\label{jzd}
\\ \nonumber
 &+&\frac{g}{2c_{W}}\sum_{i=1,6}(T^{3})_{i}\left[c^{i}_{R}\overline{Q}^{i}_{R}+s^{i}_{R}\overline{q}^{i}_{R}\right]\gamma^{\mu}\left[s^{i}_{R}q^{i}_{R}+c^{i}_{R}Q^{i}_{R}\right]-
\\ \nonumber
 &-&\frac{g}{2c_{W}}\sum_{i=1,6}(T^{3})_{i}\left[c^{i}_{L}\overline{Q}^{i}_{L}-s^{i}_{L}\overline{q}^{i}_{L}\right]\alpha\gamma^{\mu}\gamma_{5}\left[-s^{i}_{L}q^{i}_{L}+c^{i}_{L}Q^{i}_{L}\right]-
\\ \nonumber
 &-&\frac{g}{2c_{W}}\sum_{i=1,6}(T^{3})_{i}\left[c^{i}_{R}\overline{Q}^{i}_{R}+s^{i}_{R}\overline{q}^{i}_{R}\right]\alpha\gamma^{\mu}\gamma_{5}\left[s^{i}_{R}q^{i}_{R}+c^{i}_{R}Q^{i}_{R}\right].
\end{eqnarray}
Not surprisingly the couplings to the Goldstone bosons exhibit the same structure, $(\delta L_{\pi})_{\mathcal{D}}=(\delta L_{\pi})_{SM}+(\delta L_{\pi})_{2}$. Nonetheless we find convenient to write them in the following form
\begin{equation}
 (\delta L_{\pi})_{\mathcal{D}}=(\delta L_{\pi^{+}})+(\delta L_{\pi^{0}})+(\delta L_{\pi^{-}}).
\end{equation}
In the PC case we have
\begin{eqnarray}
 (\delta L_{\pi^{+}})_{PC} &=& \frac{i\sqrt{2}\pi^{+}}{v}\sum_{ij=1,3}V_{ij}\left[m_{u_{i}}c^{u_{i}}_{L}\overline{u}^{i}_{R}+M_{U_{i}}s^{u_{i}}_{L}\overline{U}^{i}_{R}\right]\left[c^{d_{j}}_{L}d^{j}_{L}+s^{d_{j}}_{L}D^{j}_{L}\right]- \label{pion}
\\ \nonumber
 &-&\frac{i\sqrt{2}\pi^{+}}{2v}\sum_{i=1,3}m^{d_{i}}_{R}\mathcal{V}_{ij}^{*}\left[-s^{u_{j}}_{L}\overline{u}^{j}_{L}+c^{u_{j}}_{L}\overline{U}^{j}_{L}\right]\left[-c^{d_{i}}_{R}d^{i}_{R}+s^{d_{i}}_{R}D^{i}_{R}\right]-
\\ \nonumber
 &-&\frac{i\sqrt{2}\pi^{+}}{2v}\sum_{ij=1,3}m^{d_{i}}_{L}\mathcal{V}_{ji}^{*}\left[s^{u_{j}}_{R}\overline{u}^{j}_{R}+c^{u_{j}}_{R}\overline{U}^{j}_{R}\right]\left[c^{d_{i}}_{L}d^{i}_{L}+s^{d_{i}}_{L}D^{i}_{L}\right]+h.c.
\\ 
 (\delta L_{\pi^{0}})_{PC} &=& \frac{i\pi^{0}}{2v}\sum_{i=1,3}\left[m_{u_{i}}c^{u_{i}}_{L}\overline{u}^{i}_{R}+M_{U_{i}}s^{u_{i}}_{L}\overline{U}^{i}_{R}\right]\left[c^{u_{i}}_{L}u^{i}_{L}+s^{u_{i}}_{L}U^{i}_{L}\right]-
\\ \nonumber
 &-&\frac{i\pi^{0}}{2v}\sum_{i=1,3}m^{u_{i}}_{R}\left[-s^{u_{i}}_{L}\overline{u}^{i}_{L}+c^{u_{i}}_{L}\overline{U}^{i}_{L}\right]\left[-c^{u_{i}}_{R}u^{i}_{R}+s^{u_{i}}_{R}U^{i}_{R}\right]-
\\ \nonumber
 &-&\frac{i\pi^{0}}{2v}\sum_{i=1,3}\left[m_{d_{i}}c^{d_{i}}_{L}\overline{d}^{i}_{R}+M_{D_{i}}s^{d_{i}}_{L}\overline{D}^{i}_{R}\right]\left[c^{u_{i}}_{L}d^{i}_{L}+s^{u_{i}}_{L}D^{i}_{L}\right]+
\\ \nonumber
&+&\frac{i\pi^{0}}{2v}\sum_{i=1,3}m^{d_{i}}_{R}\left[-s^{d_{i}}_{L}\overline{d}^{i}_{L}+c^{d_{i}}_{L}\overline{D}^{i}_{L}\right]\left[-c^{d_{i}}_{R}d^{i}_{R}+s^{d_{i}}_{R}D^{i}_{R}\right]+h.c.
\\ 
 (\delta L_{\pi^{-}})_{PC} &=&  -\frac{i\sqrt{2}\pi^{-}}{2v}\sum_{ij=1,3}m^{u_{j}}_{L}\mathcal{V}_{ij}\left[s^{d_{j}}_{R}\overline{d}^{j}_{R}+c^{d_{j}}_{R}\overline{D}^{j}_{R}\right]\left[c^{u_{i}}_{L}u^{i}_{L}+s^{u_{i}}_{L}U^{i}_{L}\right]-
\\ \nonumber
 &-&\frac{i\sqrt{2}\pi^{-}}{2v}\sum_{i=1,3}m^{u_{j}}_{R}\mathcal{V}_{ij}\left[-s^{d_{i}}_{L}\overline{d}^{i}_{L}+c^{d_{i}}_{L}\overline{D}^{i}_{L}\right]\left[-c^{u_{j}}_{R}u^{j}_{R}+s^{u_{j}}_{R}U^{j}_{R}\right]+
\\ \nonumber
 &+&\frac{i\sqrt{2}\pi^{-}}{v}\sum_{ij=1,3}V_{ji}^{*}\left[m_{d_{i}}c^{d_{i}}_{L}\overline{d}^{i}_{R}+M_{D_{i}}s^{d_{i}}_{L}\overline{D}^{i}_{R}\right]\left[c^{u_{j}}_{L}u^{j}_{L}+s^{u_{j}}_{L}U^{j}_{L}\right]+h.c.\; .
\end{eqnarray}
In the PB case, the second line of $\delta L_{\pi^{+}}$ becomes
\begin{equation}
 -\frac{i\sqrt{2}\pi^{+}}{2v}\sum_{i=1,3}m^{d_{i}}_{R}\left[-s^{u_{i}}_{L}\overline{u}^{i}_{L}+c^{u_{i}}_{L}\overline{U}^{i}_{L}\right]\left[-c^{d_{i}}_{R}d^{i}_{R}+s^{d_{i}}_{R}D^{i}_{R}\right],
\end{equation}
and also the second line of $\delta L_{\pi^{-}}$ is modified,
\begin{equation}
 -\frac{i\sqrt{2}\pi^{-}}{2v}\sum_{i=1,3}m^{u_{i}}_{R}\left[-s^{d_{i}}_{L}\overline{d}^{i}_{L}+c^{d_{i}}_{L}\overline{D}^{i}_{L}\right]\left[-c^{u_{i}}_{R}u^{i}_{R}+s^{u_{i}}_{R}U^{i}_{R}\right],
\end{equation}
while the other couplings are equal to those in the PC case.\\
The mixing between light and heavy degrees of freedom has an impact at low energies through loop effects that affect FCNC amplitudes. In addition to that composite Doublets contribute also to the EWPT \cite{Barbieri:2008zt}. 

In particular corrections to the $T$-parameter arise only after the breaking of the custodial symmetry. If the breaking is strong in the left-handed sector i.e. $m^{u}_{L}\gg m_{L}^{d}$, $\Delta T$ is always unacceptably large, whereas it can be kept under control for $m^{u}_{L}\approx m_{L}^{d}$. In this case however it is important to watch the $Z\overline{b}_{L}b_{L}$ coupling. In general the deviations from the SM value of $g_{L}$ and $g_{R}$ can be expressed as
\begin{equation}
 \delta g_{L,R}= (s_{L,R}^{b})^{2}\left[g_{L,R}(B)-g_{L,R}(b)\right],
\end{equation}
neglecting an overall factor $g/c_{W}$, from the Lagrangians of the Doublets one gets
\begin{equation}
 \delta g_{L}^{(\mathcal{D})}= \frac{(s^{b}_{L})^{2}}{4}(1-\alpha), \quad \delta g_{R}^{(\mathcal{D})}= -\frac{(s_{R}^{b})^{2}}{4}(1-\alpha).
\end{equation}
In section \ref{s4} we use this last relations to set an upper bound on the mass of the top partner. At the moment however we turn to the effect of the mixing on FCNC transitions.
\section{FCNC operators}\label{s3}
In general, it is interesting to study the new flavour structure that emerges when the composite fermions fall into fundamental representations of $SU(2)_{L+R}$. In addition to that, the high sensitivity of FCNC transitions to new physics allows to set meaningful constraints to the free parameters of the model.With these two considerations in mind, we compute the $d_{i}\to Zd_{j}$, $d_{i}\to \gamma d_{j}$ and $d_{i}\overline{d}_{j}\to d_{j}\overline{d}_{i}$  amplitudes in the $gaugeless$ limit of the model. 

In this section we are not presenting the result of the full computation, but only those terms in the amplitudes that lead to meaningful constraints, for a more complete discussion we refer to Appendix A and \cite{D'Agnolo:2009th}.

The calculation of FCNC amplitudes in the SM has been performed in \cite{Inami:1980fz} and we have adopted a similar notation.
\subsection{The Z penguin}
This is the only process, among those that we have taken into account, that, in the gaugeless limit, is affected by the value of $\alpha$ . If we take $\alpha =1$, the lowest-order chiral Lagrangian can be mapped into a version of the Standard Model with three extra generations of quarks. The transformation of the fields that does the trick is
\begin{equation}
 \psi_{L}=\frac{1-\gamma_{5}}{2}u^{\dagger}\mathcal{D}, \quad \psi_{R}=\frac{1+\gamma_{5}}{2}u\mathcal{D}
\end{equation}
and the corresponding inverse transformation reads
\begin{equation}
 \mathcal{D}=\left(u\psi_{L}+u^{\dagger}\psi_{R}\right).
\end{equation} 
Therefore the theory must be renormalizable and we expect 
the $d_{i}\to Zd_{j}$ amplitude to be finite as we have explicitly verified \cite{D'Agnolo:2009th}.
Nonetheless the key features of the result can also be discussed 
without analyzing the complete expression.
First and foremost, Lorentz and gauge invariance dictate that it can contain only the operators
\begin{eqnarray}
 (O^{\rho}_{L})_{ij}&=&\overline{d}^{j}_{L}\gamma^{\rho}d^{i}_{L}, \\ \nonumber
 (O^{\rho}_{R})_{ij}&=&\overline{d}^{j}_{R}\gamma^{\rho}d^{i}_{R}. \nonumber
\end{eqnarray}
Second, and not less important, we can easily identify the terms that 
have some phenomenological significance by their parametric 
dependence on quark masses and CKM matrix elements. To do so,
it is convenient to write the effective Lagrangian as an expansion in the Cabibbo angle $\lambda\approx 0.2$, that conveniently parametrizes also the hierarchy between down quark masses 
\begin{equation}
 \frac{m_{b}}{M_{W}}\approx \lambda^{2}, \quad \frac{m_{s}}{M_{W}}\approx \lambda^{4}, \quad \frac{m_{d}}{M_{W}}\approx \lambda^{6}.
\end{equation}
Then it is not hard to identify the dominant contribution to the amplitude, that both in the PC and PB cases, for $\alpha=1$, has the same structure of the Standard Model result. In an effective Lagrangian language, it can be written as
\begin{equation}
 \mathcal{L}_{Zij}^{T}=\frac{V_{tj}V_{ti}^{*}}{(4\pi)^{2}}\frac{g^{3}}{\cos\theta_{W}}\left[(c_{L}^{t})^{2}\Gamma(x_{t})+(s_{L}^{t})^{2}\Gamma(x_{T})\right](O_{L}^{\rho})_{ij}Z_{\rho}, \label{Lzt}
\end{equation}
where
\begin{equation}
 x_{t}=\left(\frac{m_{t}}{M_{W}}\right)^{2}, \quad x_{T}=\left(\frac{M_{T}}{M_{W}}\right)^{2}
\end{equation}
and $M_{T}$ is the physical mass of the top partner. The function $\Gamma$ in (\ref{Lzt}) is the usual Inami-Lim function
\begin{equation}
 \Gamma(x)=\frac{x}{4}-\frac{1}{2}(Q\sin^{2}\theta_{W}+1)\left[\frac{x^{2}}{(x-1)^{2}}\log x-\frac{x}{(x-1)}\right], \quad Q=-\frac{1}{3}
\end{equation}
computed for those diagrams where there are not propagating gauge bosons.

Strictly speaking, the gaugeless limit implies a vanishing $Z$-penguin
amplitude. What we have considered here is a modified version of this limit, namely 
an expansion of the amplitude in the gauge couplings up to $O(g)$. At this order,
where only Goldstone bosons and fermions contribute to the amplitude, we find
\begin{equation}
\Gamma(x_{i})\longrightarrow \frac{x_{i}}{4}, \quad i=t,T.
\end{equation}
This is all that we need to constrain the two free parameters in the top sector, as we will 
see in section \ref{s4}. The second set of bounds that we would like to find is on the matrix elements of $\mathcal{V}$. The Z penguin significantly participates in setting these limits only in the PB case, through the effective Lagrangian
\begin{equation}
  \mathcal{L}_{Zij}^{\mathcal{V}}=\frac{\mathcal{V}_{tj}\mathcal{V}_{ti}^{*}}{(4\pi)^{2}}\frac{g^{3}}{\cos\theta_{W}}\frac{s_{R}^{i}s_{R}^{j}}{2}(m_{L}^{t})^{2}\left[(c_{L}^{t})^{2}\frac{\Delta(x_{t})}{m_{t}^{2}}+(s_{L}^{t})^{2}\frac{\Delta(x_{T})}{M_{T}^{2}}\right](O_{R}^{\rho})_{ij}Z_{\rho}, \label{Lzv}
\end{equation}
where, again we have considered the case $\alpha=1$ 
and the loop function is 
\begin{equation}
 \Delta(x)=\frac{x}{4}+\frac{1}{2}(-Q\sin^{2}\theta_{W}+\frac{1}{4})\left[\frac{x^{2}}{(x-1)^{2}}\log x-\frac{x}{(x-1)}\right].
\end{equation}
which in the gaugeless limit reduces to
\begin{equation}
 \Delta(x_{i})\longrightarrow \frac{x_{i}}{4}, \quad i=t,T.
\end{equation}

 For $\alpha \not=1$ the flavour structure of the 
two effective Lagrangians in (\ref{Lzt}) and (\ref{Lzv}) is the same; however, 
the one-loop calculation is divergent~\cite{D'Agnolo:2009th} and we
need to include appropriate counterterms to regularize it. 
This results in unknown subleading corrections
of order $(\alpha-1)$ to both the $\Gamma$ and $\Delta$ 
functions.

\subsection{The $\Delta F=2$ transitions}
The operators that appear in the effective Lagrangian are
\begin{equation}
 \begin{array}{rcl}
 \mathcal{O}_{LL}&=&(\overline{s}_{L}\gamma_{\rho}b_{L})^{2}\\ \nonumber
 \mathcal{O}_{LR}&=&(\overline{s}_{L}\gamma_{\rho}b_{L})(\overline{s}_{R}\gamma^{\rho}b_{R}) \\ \nonumber  
\mathcal{O}_{RR}&=&(\overline{s}_{R}\gamma_{\rho}b_{R})^{2}
\end{array}
\qquad
\begin{array}{rcl}
\mathcal{O}^{'}_{L}&=&(\overline{s}_{R}b_{L})^{2} \\ \nonumber  \mathcal{O}^{'}_{R}&=&(\overline{s}_{L}b_{R})^{2} \\ \nonumber \mathcal{O}^{'}_{LR}&=&(\overline{s}_{R}b_{L})(\overline{s}_{L}b_{R}),
\end{array}
\end{equation}
where again the transition involves $b$ and $s$ quarks and can be easily generalized. These operators are present in the result of the full computation. However the only one we have used to set limits on the observables is $\mathcal{O}^{'}_{L}$ due to the interplay between renormalization group enhancement~\cite{Davidson:2007si, Buras:2001ra} and relative importance 
of the coefficients arising from the couplings to Goldstone bosons.
Therefore we can write the relevant part of the amplitude, both in the PB and PC cases, as
\begin{eqnarray}
\label{box}
 \mathcal{L}_{ij}^{box}&=&\frac{g^{4}}{64\pi^{2}M_{W}^{6}}\left(\frac{\mathcal{V}_{tj}V_{ti}}{2}s^{j}_{R}c^{i}_{L}m_{L}^{t}\right)^{2}\left[(c_{L}^{t})^{4}m^{2}_{t}G(x_{t}, x_{t})+ \right.\\\nonumber
&+&\left.2(c_{L}^{t}s_{L}^{t})^{2}m_{t}M_{T}G(x_{t}, x_{T})+(s_{L}^{t})^{4}M^{2}_{T}G(x_{T}, x_{T})\right]\mathcal{O}^{'}_{L}. 
\end{eqnarray}
We have introduced a new loop function, identical to one of those that appear in the Standard Model computation, which is
\begin{eqnarray}
 G(x,y) &=& -\frac{1}{8}\frac{1}{y-x}\left[\frac{y}{(y-1)^{2}}\log y - \frac{x}{(x-1)^{2}}\log x-\frac{1}{y-1}+\frac{1}{x-1}\right], \\
 G(x,x) &=& \lim_{y\rightarrow x}G(x,y)=\frac{2-2x+(1+x)\log x}{(x-1)^{3}}.
\end{eqnarray}
For this transition we can take an actual gaugeless limit without sending the amplitude of the process to zero. As $g\rightarrow 0$ we have
\begin{eqnarray}
 \frac{G_{1}(x,y)}{M_{W}^{4}}&\longrightarrow& \frac{1}{8}\frac{1}{m_{x}^{2}-m_{y}^{2}}\left[\frac{1}{m_{y}^{2}}\log \frac{m_{y}^{2}}{M_{W}^{2}}-\frac{1}{m_{x}^{2}}\log \frac{m_{x}^{2}}{M_{W}^{2}}\right], \\
 \frac{G_{1}(x,x)}{M_{W}^{4}}&\longrightarrow& \log \frac{m_{x}^{2}}{M_{W}^{2}}.
\end{eqnarray}
\subsection{The photon penguin}
We can now turn to the $d_{j}\gamma d_{i}$ vertex. We are interested in the operator $O^{1}_{ij}=\overline{d}^{j}_{L}\sigma_{\mu\nu}F^{\mu\nu}d^{i}_{R}$ (or equivalently $O^{2}_{ij}=\overline{d}^{j}_{R}\sigma_{\mu\nu}F^{\mu\nu}d^{i}_{L}$). With this in mind and gauge invariance on our side we can compute the corresponding effective Lagrangian
\begin{eqnarray}
\label{photon}
 \mathcal{L}_{ij}^{\gamma}=\frac{e (Q+1)}{(4\pi)^{2}}\frac{g^{2}}{2M_{W}^{2}}\frac{\mathcal{V}_{tj}V_{ti}}{2}s^{j}_{R}c^{i}_{L}m_{L}^{t}\left[(c_{L}^{t})^{2}\frac{A_{SM}(x_{t})}{m_{t}}+(s_{L}^{t})^{2}\frac{A_{SM}(x_{T})}{M_{T}}\right]O^{2}_{ij}, \quad Q=-\frac{1}{3}
\end{eqnarray}
Also in this case the loop function is the same appearing 
in the  Standard Model, whose explicit expression can be 
found in~\cite{Inami:1980fz}.
%
\section{Experimental limits}\label{s4}
This work has a twofold aim. In addition to presenting the computation of the FCNC Lagrangians in a rather generic framework, that, being a complete result, is interesting in itself, we have been able to constrain meaningfully the Higgsless models of EWSB in which the mass generation proceeds through mixing. In this section we give the results related to this second respect of the work. In order to set bounds on the free parameters of the theory, we have used the model independent analysis of $\Delta F=2$ effective Hamiltonians performed in \cite{Bona:2007vi} and the results on $\Delta F=1$ transitions in \cite{Davidson:2007si}. For simplicity, from now on, we drop 
the subscript $t$ in $M_{\mathcal{D}_{t}}$, writing $M_{\mathcal{D}}$ instead.
\subsection{Massless down quarks}
In the limit of massless down quarks the leading contributions to FCNC observables survive. Since neglecting terms of order $\frac{m_{d_{i}}}{M_{W}}$ can not shift the bounds set in this section significantly, we proceed examining first this limiting case.
\subsubsection{The top sector}
\paragraph{FCNC transitions in the PC case.}
The first natural step is to constrain the two free parameters in the top sector, which is the only one where the light-heavy mixing can considerably affect FCNC transitions, as noted in section \ref{s23}.

\begin{figure}[t]
  \centering%
  \subfigure[\label{fig: 1aB}]%
    {\includegraphics[scale=0.445]{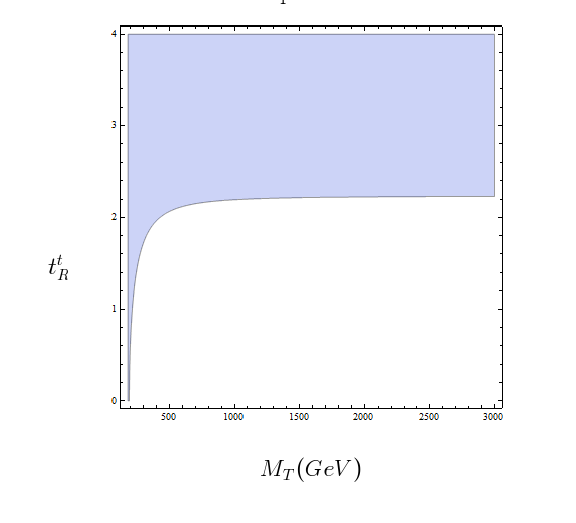}}\qquad\qquad
  \subfigure[\label{fig: 1bB}]%
    {\includegraphics[scale=0.5]{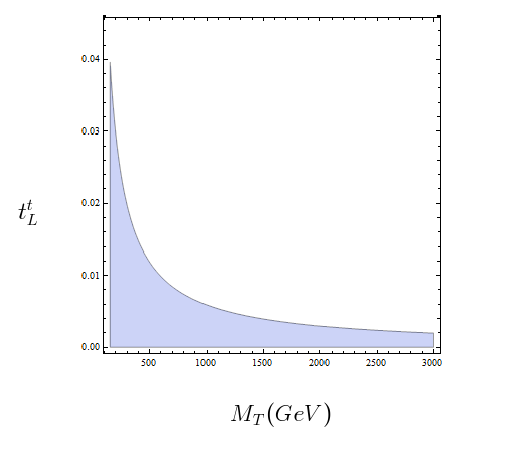}}\qquad\qquad
  \caption{Bounds at $95\%$ C.L. from $d_{i}\to Z d_{j}$ with $\alpha=1$ in the $t_{R}^{t}-M_{T}$ and  $t_{L}^{t}-M_{T}$ planes, for massless down quarks. The allowed region is the cerulean one.}
\label{fig: 1B}
\end{figure}

In Figure \ref{fig: 1B} we have the bound at $95\%$ C.L. in the $t_{R}^{t}-M_{T}$ and $t_{L}^{t}-M_{T}$  planes, arising from the Z penguin (\ref{Lzt}). The allowed region, coloured in light blue, has been obtained imposing
\begin{equation}
 \left|\frac{\mathcal{A}_{SM}-\mathcal{A}_{\mathcal{D}}}{\mathcal{A}_{SM}}\right|< c, \quad c\approx \frac{1}{5} \label{limit}
\end{equation}
since 
\begin{equation}
 \mathcal{A}_{\mathcal{D}}(x_{t}, x_{T})= c^{2}_{L}\mathcal{A}_{SM}(x_{t})+s^{2}_{L}\mathcal{A}_{SM}(x_{T}). \label{AD}
\end{equation}
With $\mathcal{A}_{\mathcal{D}}$ we have indicated the main contribution to the amplitude for the process in the Composite Fermions model, coming from (\ref{Lzt}).

 The amplitude $A_{\mathcal{D}}$ contains the \textit{centered} SM value (the only one left in the limit $s_{L}\to 0$) plus a correction, as can be seen in (\ref{AD}). It is only this correction that should not exceed approximately $20\%$ \cite{Davidson:2007si, Bona:2007vi} of $\mathcal{A}_{SM}$.
The shape of the allowed region can be explained as follows
\begin{eqnarray}
 t_{R}&\to& \infty \Rightarrow s_{L}\approx 0 \Rightarrow \mathcal{A}_{\mathcal{D}}\approx \mathcal{A}_{SM}(x_{t})\Rightarrow \left|1-\frac{\mathcal{A}_{\mathcal{D}}}{\mathcal{A}_{SM}}\right|\approx 0 \\
M_{T}&\approx& m_{t} \Rightarrow \mathcal{A}_{\mathcal{D}}\approx \mathcal{A}_{SM}(x_{t})\Rightarrow \left|1-\frac{\mathcal{A}_{\mathcal{D}}}{\mathcal{A}_{SM}}\right|\approx 0\label{mtl} \\
M_{T}&\to& \infty \Rightarrow \mathcal{A}_{\mathcal{D}}\approx s_{L}^{2}\frac{M_{T}^{2}}{m_{t}^{2}}\approx \frac{M_{T}^{2}}{f_{t}^{2}} \label{mf}
\end{eqnarray}
or simply by noticing that the relevant part of the amplitude can be written, in the gaugeless limit, as
\begin{equation}
 \mathcal{A}_{\mathcal{D}}\propto \frac{1}{1+\left(\frac{m_{t}}{t_{R}^{t}M_{T}}\right)^{2}}\left[1+\frac{1}{(t^{t}_{R})^{2}}\right] \quad \mathrm{or} \quad \mathcal{A}_{\mathcal{D}}\propto \frac{1}{1+(t_{L}^{t})^{2}}\left[1+\left(\frac{M_{T}t_{L}^{t}}{m_{t}}\right)^{2}\right]
\end{equation}
Where we have used $t^{q}_{L}t^{q}_{R}=m_{q}/M_{Q}$.\\
The ratio $M_{T}/f_{t}$, in (\ref{mf}), is constant if we take the limit $M_{T}\to \infty$ with $m_{t}$ fixed (or equivalently with $t^{t}_{R}$ fixed), this explains why for $M_{T}\to \infty$ the area of the allowed region does not grow. 
In addition to that the correction to the SM value vanishes for $M_{T}=m_{t}$, as expected in the case $\alpha=1$.

We are confronted with a new phenomenon. In the case of the Singlets, in fact, some diagrams had also the decoupling limit $M_{T}\gg m_{t}$ \cite{Barbieri:2008zt}, but this is not possible for Doublets, because of the different nature of the mixing Lagrangian, where we have Yukawa couplings to Goldstone bosons, arising also from $m_{R}^{q}$, that were not present in the Singlets case.

\begin{figure}[t]
 \centering
 \includegraphics[bb=0 0 424 370, scale=0.6]{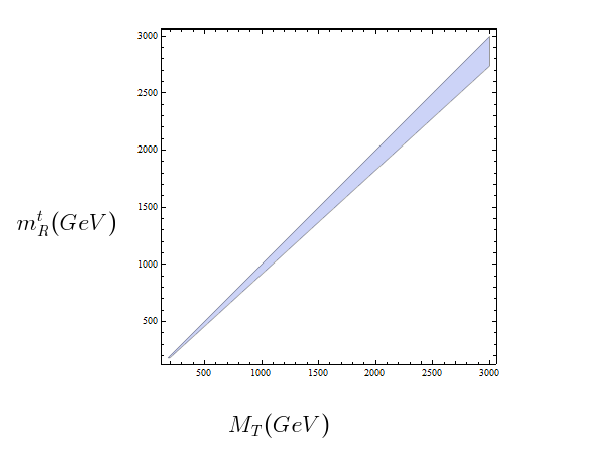}
 \caption{Bound at $95\%$ C.L. from $d_{i}\to Z d_{j}$ with $\alpha=1$ in the $m_{R}^{t}-M_{T}$ plane, for massless down quarks. The allowed region is the cerulean one.}
 \label{fig: AB}
\end{figure}

We could obtain the same decoupling effect present in the SM and the Singlets case, sending $t_{R}^{t}~\to ~\infty$, but this would create infinite FCNC amplitudes associated with operators that otherwise do not lead to significant constraints. This can be easily seen by direct inspection of the full computation of the amplitudes presented in Appendix A. Any operator with a coupling proportional to $m_{R}^{t}$, in fact, would give a diverging contribution.\\
This is not the whole story, though, since the relation between $t_{R}^{t}$ and $M_{T}$
\begin{equation}
 t_{R}^{t}=\frac{(m^{t}_{R})^{2}-(m^{t}_{L})^{2}-M_{\mathcal{D}}^{2}+\sqrt{M_{\mathcal{D}}^{4}+2M_{\mathcal{D}}^{2}((m_{L}^{t})^{2}+(m_{R}^{t})^{2})+((m_{L}^{t})^{2}-(m_{R}^{t})^{2})^{2}}}{2m_{R}M_{\mathcal{D}}},
\end{equation}
\begin{equation}
 t_{R}^{t}\approx \frac{m_{R}^{t}}{\sqrt{M^{2}_{T}-(m^{t}_{R})^{2}}}\quad \mathrm{for}\; M_{\mathcal{D}}\gg m^{t}_{L} \label{tmr}
\end{equation}
hides the fact that the allowed region in the $t_{R}^{t}-M_{T}$ plane is largely non-natural. This is manifest if we repeat the same exercise in the $m_{R}^{t}-M_{T}$ plane. The result can be seen in Figure \ref{fig: AB}.\\
 This graph is far more instructive than the previous one. First of all, the decoupling limit of the left-handed sector, obtained for $m_{R}^{t}\to \infty$, is displayed here as a function of a parameter in the Lagrangian. Second and not less important, it tells us that solely for $M_{T}$ very close to $m_{R}^{t}$ is possible to have sufficiently small FCNC amplitudes in the left-handed sector. The slice of parameter space left open is just $0.9\times m_{R}^{t}\lesssim M_{T} < m_{R}^{t}$.

The profile of the allowed region can be described partly by the simple relation (\ref{tmr}), which favours a cone around the line $m_{R}^{t}=M_{T}$ that broadens as $m_{R}^{t}$ increases. Nonetheless, this equation can not account for the fact that the region in which $m_{R}^{t}> M_{T}$ is forbidden. This is due to the requirement that the mass parameter in the Doublets Lagrangian, $M_{\mathcal{D}}$, be real.

Another advantage of this plot is that $m_{R}^{t}$, having dimension of mass, has a natural cut-off, that has no direct influence on $t_{R}^{t}$. 

\begin{figure}[htp]
  \centering%
  \subfigure[\label{fig: BB}]%
    {\includegraphics[scale=0.45]{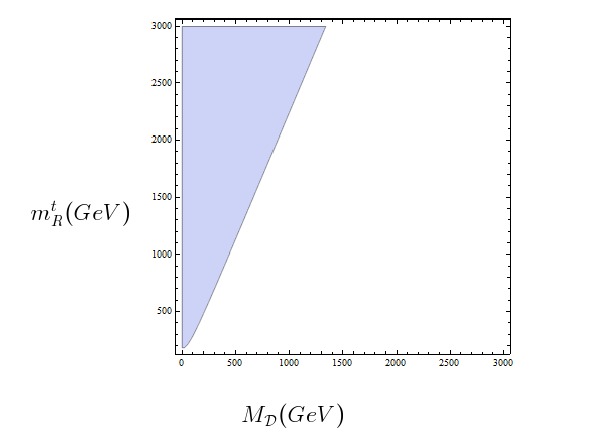}}\qquad\qquad
  \subfigure[\label{fig: CB}]%
    {\includegraphics[scale=0.45]{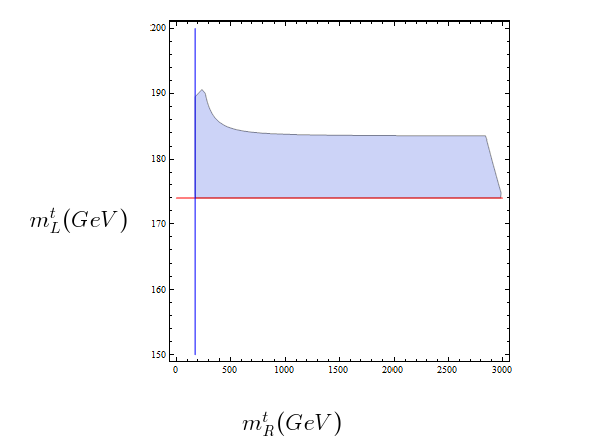}}\qquad\qquad
  \caption{Bounds at $95\%$ C.L. from $d_{i}\to Z d_{j}$ with $\alpha=1$ in the (a) $m_{R}^{t}-M_{\mathcal{D}}$ and (b) $m_{L}^{t}-m_{R}^{t}$ planes, for massless down quarks. The allowed region is the cerulean one. In (b) the red line corresponds to $m_{L}^{t}=m_{t}$, while the blue line to $m_{R}^{t}=m_{t}$.}
\label{fig: B}
\end{figure}

It is also appropriate to study the relation between the non-physical parameters of the Lagrangian, since Figure \ref{fig: AB} implies a certain amount of fine tuning among them. This intuition is readily confirmed both in the $m_{R}^{t}-M_{\mathcal{D}}$ plane and in the $m_{L}^{t}-m_{R}^{t}$ plane, as can be seen in Figures \ref{fig: BB} and \ref{fig: CB}.
In the first of the two plots we see that $M_{\mathcal{D}}\lesssim 2 m_{R}^{t}$, never exceeding 1.5 TeV. This is a serious constraint that can be met only tuning the parameters of the model. There is no possible symmetry that bounds $M_{\mathcal{D}}$ to live in the allowed region, unless one is interested in a model with $M_{\mathcal{D}}=0$, which would have quite a different phenomenology from the class of theories that have the Composite Fermions model as their low energy limit.

\begin{figure}[htp]
  \centering%
  \subfigure[\label{fig: alfaaB}]%
    {\includegraphics[scale=0.5]{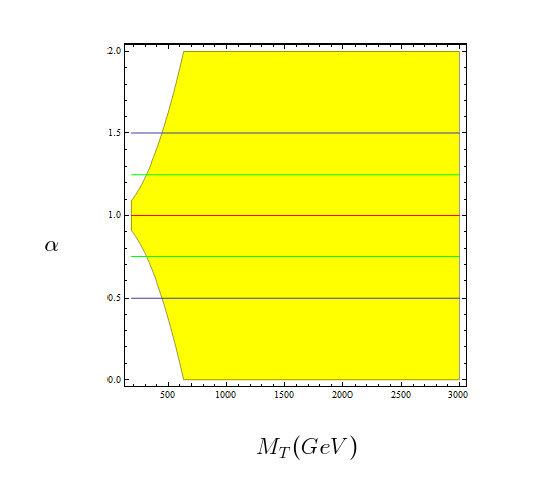}}\qquad\qquad
  \subfigure[\label{fig: alfabB}]%
    {\includegraphics[scale=0.5]{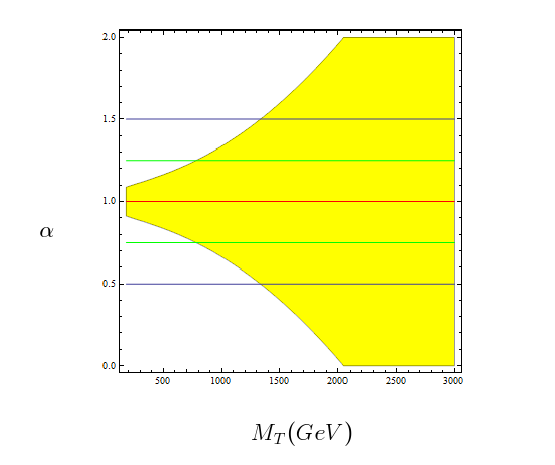}}\qquad\qquad
  \subfigure[\label{fig: alfacB}]%
    {\includegraphics[scale=0.5]{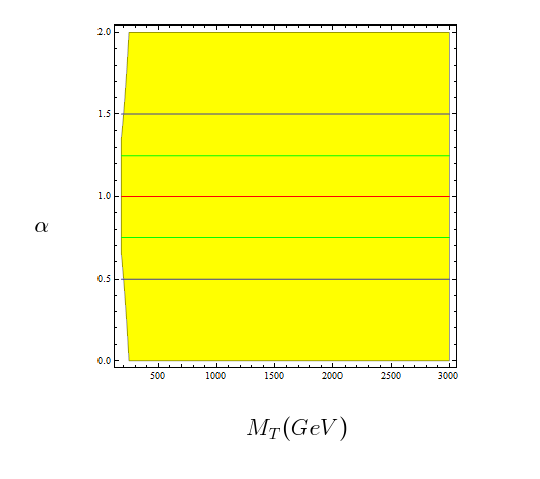}}
  \caption{The constraint at 95$\%$ confidence level in the $\alpha-M_{T}$ plane, arising from the limit on $\delta g_{L}$ obtained by the measurement of $R_{b}$ in \cite{:2005ema}. The red line indicates $\alpha=1$. Inside the blue band $\frac{\Delta \alpha}{\alpha}< 100 \%$, while inside the green band $\frac{\Delta \alpha}{\alpha}< 50 \%$. The allowed region is coloured in yellow. The three different plots correspond to (a) $t_{R}^{t}=2.5$ and $t_{R}^{b}=2.5$, (b) $t_{R}^{t}=10$ and $t_{R}^{b}=2.5$, (c) $t_{R}^{t}=2.5$ and $t_{R}^{b}=5$.\label{fig: alfaB}}
\end{figure}

In Figure \ref{fig: CB} the fine tuning is even more evident, since $m_{L}^{t}$ is relegated to a narrow band of the parameter space, that is $m_{t}< m_{L}^{t}\lesssim 195$ GeV. 
\paragraph{The $Zb\overline{b}$ vertex.}
The FCNC amplitudes with the exchange of a top quark in the loop, are not the only source of fine-tuning the model has to cope with, another constraint, that involves the top sector, arises from the experimental limits on $\delta g_{L}$ described in section \ref{s24}. This limit is depicted in Figure \ref{fig: alfaB} for different values of $t_{R}^{t}$ and $t_{R}^{b}$. We have chosen such values in order to be able to vary freely $M_{T}$ without violating the bounds set by FCNC transitions.

If $t_{R}^{b}=2.5$, we are able to extract a lower bound on $M_{T}$ fixing the amount of fine-tuning on $\alpha$ that we consider acceptable.
For instance $\frac{\Delta \alpha}{\alpha}> 50 \%$ corresponds to $M_{T}\gtrsim312$ GeV if $t_{R}^{t}=2.5$ while $M_{T}\gtrsim 780$ GeV if $t_{R}^{t}=10$. However, as can be seen in Figure \ref{fig: alfacB}, the bound evaporates as soon as $s_{L}^{b}$ decreases. This is in perfect agreement with intuition, since as the mixing weakens in the bottom sector, the $Zb_{L}\overline{b}_{L}$ vertex resembles the more and more its SM counterpart. It might be argued that, for large values of $t_{R}^{t}$, $\delta g_{R}$ starts playing an important role. Nonetheless it is a straightforward exercise to prove that it is always negligible with respect to the SM value, regardless of the behaviour of $s_{R}^{t}$. 
\subsubsection{The matrix elements $\mathcal{V}_{ij}$}
In the PB case we can constrain the CKM-like matrix elements $\mathcal{V}_{ij}$, already in the limit of massless down quarks. To this end, we first notice that the bounds on $M_{T}$ and $t_{R}^{t}$ are the same already obtained in the PC case, since the contributions of left-handed operators to the amplitudes, corresponding to the two breaking patterns of the flavour symmetry, differ only by terms that are suppressed, with respect to the leading one, by powers of $m_{L}^{d}/M_{W}$. In addition to that,  including also the new right-handed operators that appear in the massless PB case, does not lead to significant modifications of the bounds discussed above, if we take the $\mathcal{V}_{ij}$'s to be equal or smaller than the CKM matrix elements. In the following, this turns out to be a very reasonable assumption. 

\begin{figure}[t]
  \centering%
  \subfigure[\label{fig: 3aB}]%
    {\includegraphics[scale=0.5]{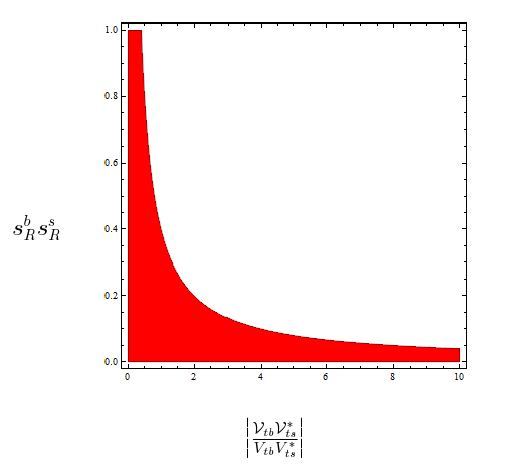}}\qquad\qquad
  \subfigure[\label{fig: 3bB}]%
    {\includegraphics[scale=0.5]{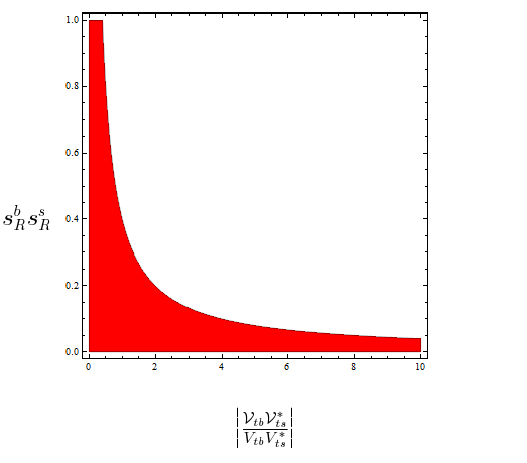}}
  \caption{Bounds in the plane $\mathcal{V}^{2}-s^{b}_{R}s^{s}_{R}$ from the Z penguin with $\alpha=1$ (PB case), for different values of $t_{L}$ and $M_{T}$. We have respectively (a) $t_{L}=0.001$, $M_{T}=3$ TeV, (b) $t_{L}=0.004$, $M_{T}=1$ TeV. The allowed region is red.\label{fig: 3B}} 
\end{figure}

In Figure \ref{fig: 3B} the limits in the plane $\mathcal{V}^{2}-s^{b}_{R}s^{s}_{R}$ are depicted for the Z penguin (\ref{Lzv}) in the case $\alpha=1$. The constraints from the $\Delta F=2$ transitions are less stringent. \\
Unfortunately it is not possible to constrain the $\mathcal{V}_{ij}$'s alone, since every amplitude has the form
\begin{equation}
 \mathcal{A}_{\mathcal{D}}\propto \mathcal{V}M(\lambda_{d_{i}}),
\end{equation}
where $M(\lambda_{d_{i}})$ indicates a generic combination of mixing angles and masses of the composite partners in the down quarks sector. Therefore the best we can get is an hyperbola in the plane $\mathcal{V}-M(\lambda_{d_{i}})$.\\
However we can still say something on the $\mathcal{V}_{ij}$'s. Obviously, the greater they are, the more fine tuning is needed in the down quarks sector. 

We can see that if the new angles are aligned with the CKM ones, or smaller, almost any value for the sines of the two mixing angles is allowed, on the contrary we need some amount of fine tuning if the $\mathcal{V}_{ij}$'s grow. Even for an approximately decoupled top partner, as it is the case in Figure \ref{fig: 3aB}, we can have angles greater than the CKM ones of one order of magnitude, only for very small values of the right mixing angles of the down quarks.

 In addition to that, it is worth noticing that the case $\mathcal{V}=0(1)$ is not excluded, even if it can not be seen in the figures, provided that the product $s^{b}_{R}s^{s}_{R}$ is small enough.

To make the above considerations more quantitative, we give a direct estimate of the amount of fine tuning needed for a fixed value of the $\mathcal{V}_{ij}$'s:
\begin{equation}
 s^{s}_{R}s_{R}^{b}\lesssim 0.4 \left|\frac{V_{ts}^{*}V_{tb}}{\mathcal{V}_{ts}^{*}\mathcal{V}_{tb}}\right|, \quad \mathrm{for} \;\; t_{L}^{t}=0.004, \; M_{T}=1 \;\mathrm{TeV}. \label{est}
\end{equation}
This bound is fairly insensitive to the chosen values of $t_{L}^{t}$ and $M_{T}$, provided that they are in the allowed region. For instance, if $t_{L}=0.006$, $M_{T}=600$~GeV or $t_{L}=0.004$, $M_{T}=1$~TeV, it does not differ appreciably from (\ref{est}).

 In this section we have presented the results only for the $b\to s$ transition, but, if we take into account only the top exchange in the loop, analogous considerations hold for any pair of $d_{i}$'s. This assumption seems in contrast with the intuition coming from the SM, but it is justified by the fact that the only significant impact of the mixing is in the top sector, as pointed out in section~\ref{s23}.
\subsection{Massive down quarks}
In this section we derive bounds for the matrix elements $\mathcal{V}_{ij}$ in the PC case. We have presented limits on the same matrix element arising from different processes, even if one of the alternatives gave a weaker constraint. This is due to the fact that we can have only a simultaneous limit on the $\mathcal{V}_{ij}$ and the mixing angles in the down sector, that are not constrained otherwise. Therefore we have chosen to show the most significant bound in each plane.
\subsubsection{$b\to s$}
The relevant amplitude is the photon penguin (\ref{photon}). The bound has been obtained just like in the PB case and the same considerations apply.
Even the limits are very similar to those already seen for the PB case, we have in fact,
\begin{equation}
 s^{s}_{R}c_{L}^{b}\lesssim 0.4 \left|\frac{V_{ts}^{*}}{\mathcal{V}_{ts}^{*}}\right|, \quad \mathrm{for} \;\; t_{L}^{t}=0.004, \; M_{T}=1 \;\mathrm{TeV}.
\end{equation}
\subsubsection{$b\to d$}
The relevant amplitude is the $\Delta F=2$ transition (\ref{box}). Again the allowed region shows the same structure already described for the PB case. Here the bounds are similar to the previous ones and, again, prescribe a good alignment between CKM matrix elements and the new angles $\mathcal{V}_{ij}$.
This is easily seen thanks to the usual estimate
\begin{equation}
 s^{d}_{R}c_{L}^{b}\lesssim 0.2 \left|\frac{V_{td}^{*}}{\mathcal{V}_{td}^{*}}\right|, \quad \mathrm{for} \;\; t_{L}^{t}=0.004, \; M_{T}=1 \;\mathrm{TeV}.
\end{equation}
\subsection{$d\to s$}
Once more the relevant amplitude is the $\Delta F=2$ transition (\ref{box}). The renormalization group enhancement of operators, like $\mathcal{O}'_{L}$, make this limit by far the most stringent.
\begin{equation}
 s^{d}_{R}c_{L}^{s}\lesssim 0.04 \left|\frac{V_{td}^{*}}{\mathcal{V}_{td}^{*}}\right|, \quad \mathrm{for} \;\; t_{L}^{t}=0.004, \; M_{T}=1 \; \mathrm{TeV}.
\end{equation}
Even the alignment of the new angles with the CKM ones would not allow to account for experimental data without a good amount of fine tuning.
\section{Conclusion}
Using the present experimental knowledge of FCNC transitions
 we have been able to significantly constrain the free parameters
of the model. The results obtained can be summarized as follows:
\begin{itemize}
\item[1] Bounds on mass and mixing of the top partner.

From the analysis of the FCNC Z-penguin amplitude we have not been able to set stringent limits on the mass of the top partner (see Figure \ref{fig: AB}), however the data point to a very narrow slice of the parameters space. This considerably reduces the variety of possible models in which fermions acquire masses through mixing. In view of the nature of the Composite Fermions model this is as significant as an upper limit on the mass of the composite top partner, since it allows us to exclude the theories that live outside the small window of phase space left open.\\
In addition to that we have been able to constrain the mixing angle 
in the top sector, which must be small (see Figure \ref{fig: 1B}) 
and to set a lower bound on $M_{T}$, for certain values of $t_{R}^{b}$, 
with naturalness arguments.

 ×\item[2] Bounds on the new mixing matrix $\mathcal{V}$.

The second task that confronted us was to set limits on the inter-generation mixing matrix $\mathcal{V}$. Not surprisingly the data favour matrix elements aligned with the CKM ones. Another possibility, that would trivially be consistent with the limits, is to take $\mathcal{V}$ proportional to the identity matrix. Therefore it is legitimate to ask whether one of these two conditions can be deduced from symmetry principles. The answer we found is not fully satisfying, since we could not think of any obvious symmetry doing the trick in either of the two cases.\\
The only option left is an accidental (or dynamical) alignment, in flavour space, of the two parameters $Y_{u}$ and $Y_{d}$. 
\end{itemize}

\section*{Acknowledgements}
I would like to thank G. Isidori for his support, insightful comments and the careful reading of the paper and G. Martinelli for very helpful discussions and his support.
\appendix
\section{Appendix}
In this appendix we give the result of the full computation, specialized to the case $d_{i}=b$, $d_{j}=s$. 
\subsection{Useful definitions}
We begin by introducing the definitions
\begin{eqnarray}
 a_{i,j}^{E} & = & -\mathcal{V}_{ji}s^{i}_{R}c_{L}^{j}\frac{m^{u_{j}}_{L}}{\sqrt{2}v}-V_{ji}^{*}c^{i}_{L}c_{L}^{j}\frac{\sqrt{2}m_{i}}{v}, \label{def}\\ \nonumber
 a_{i,j}^{C} & = & -\mathcal{V}_{ji}s^{i}_{R}s_{L}^{j}\frac{m^{u_{j}}_{L}}{\sqrt{2}v}-V_{ji}^{*}c^{i}_{L}s_{L}^{j}\frac{\sqrt{2}M_{i}}{v}, \\ \nonumber
 b_{i,j}^{E} & = & -\mathcal{V}_{ij}s^{i}_{L}c_{R}^{j}\frac{m^{u_{j}}_{R}}{\sqrt{2}v}[s^{i}_{L}c_{R}^{j}\frac{m^{u_{j}}_{R}}{\sqrt{2}v}], \\ \nonumber
 b_{i,j}^{C} & = & \mathcal{V}_{ij}s^{i}_{L}s_{R}^{j}\frac{m^{u_{j}}_{R}}{\sqrt{2}v}[s^{i}_{L}s_{R}^{j}\frac{m^{u_{j}}_{R}}{\sqrt{2}v}], \\ \nonumber
 c_{i,j}^{E} & = & -V_{ji}c^{i}_{L}c_{L}^{j}\frac{\sqrt{2}m_{j}}{v}+\mathcal{V}_{ji}^{*}c^{i}_{L}s_{R}^{j}\frac{m^{d_{i}}_{L}}{\sqrt{2}v}, \\ \nonumber
 c_{i,j}^{C} & = & -V_{ji}c^{i}_{L}s_{L}^{j}\frac{\sqrt{2}M_{j}}{v}+c^{i}_{L}c_{R}^{j}\mathcal{V}_{ji}^{*}\frac{m^{d_{i}}_{L}}{\sqrt{2}v}, \\ \nonumber
d_{i,j}^{E} & = & \mathcal{V}_{ij}^{*}c_{R}^{i}s_{L}^{j}\frac{m^{d_{i}}_{R}}{\sqrt{2}v}[c_{R}^{i}s_{L}^{j}\frac{m^{d_{i}}_{R}}{\sqrt{2}v}], \\ \nonumber
d_{i,j}^{C} & = & \mathcal{V}_{ij}^{*}c_{R}^{i}c_{L}^{j}\frac{m^{d_{i}}_{R}}{\sqrt{2}v}[c_{R}^{i}s_{L}^{j}\frac{m^{d_{i}}_{R}}{\sqrt{2}v}],
\end{eqnarray}
which come directly from the structure of the couplings of fermions to Goldstone bosons (\ref{pion}). The values within square brackets refer to the PB case. The definitions without alternatives in square brackets are the same for both the PC and the PB case.

Here the first index runs over all down type quarks, $i=d,s,b$, whereas the second index encompasses all the families of up type quarks $j=u,c,t$. The labels $E=Elementary$ and $C=Composite$ allow to identify the nature of the quark circulating in the loop.\\
We find convenient to introduce also a second set of definitions:
\paragraph{PC case}   
$$
\begin{array}{rcl}
 \alpha_{jR}^{E,C} &=&a_{s,j}^{E,C}(a_{b,j}^{E,C})^{*},\\ 
 \alpha_{jL}^{E,C}&=&(c_{s,j}^{E,C})^{*}c_{b,j}^{E,C},\\ 
 \beta_{jR}^{E,C}& = &(c_{s,j}^{E,C}a_{b,j}^{E,C})^{*},\\ 
 \beta_{jL}^{E,C}& = &a_{s,j}^{E,C}c_{b,j}^{E,C},\\ 
 \gamma_{jL}^{E,C}& = &(c_{s,j}^{E,C})^{*}b_{b,j}^{E,C},\\ 
 \gamma_{jR}^{E,C}& = &c_{b,j}^{E,C}(b_{s,j}^{E,C})^{*},\\ 
\delta_{jL}^{E,C}& = &a_{s,j}^{E,C}b_{b,j}^{E,C},\\ 
 \delta_{jR}^{E,C}& = &(a_{b,j}^{E,C})^{*}(b_{s,j}^{E,C})^{*},
\end{array}
\qquad 
\begin{array}{rcl}
 \epsilon_{jR}^{E,C}& = &(c_{s,j}^{E,C})^{*}(d_{b,j}^{E,C})^{*}, \\ 
 \epsilon_{jL}^{E,C}& = &c_{b,j}^{E,C}d_{s,j}^{E,C}, \\ 
 \zeta_{jR}^{E,C}& = &a_{s,j}^{E,C}(d_{b,j}^{E,C})^{*}, \\ 
 \zeta_{jL}^{E,C}& = &(a_{b,j}^{E^{E,C},C})^{*}d_{s,j}^{E,C}, \\ 
\eta_{jL}&=& b_{b,j}^{E,C}(b_{s,j}^{E,C})^{*}, \\ 
\eta_{jR}&=& (d_{b,j}^{E,C})^{*}d_{s,j}^{E,C}, \\ 
\theta_{jL}&=& b_{b,j}^{E,C}d_{s,j}^{E,C}, \\ 
\theta_{jR}&=& (d_{b,j}^{E,C})^{*}(b_{s,j}^{E,C})^{*}.
\end{array}
$$
\paragraph{PB case}
$$
\begin{array}{rcl}
 \alpha_{jR}^{E,C} &=&a_{s,j}^{E,C}(a_{b,j}^{E,C})^{*},\\ 
 \alpha_{jL}^{E,C}&=&(c_{s,j}^{E,C})^{*}c_{b,j}^{E,C},\\ 
 \beta_{jR}^{E,C}& = &(c_{s,j}^{E,C}a_{b,j}^{E,C})^{*},\\ 
 \beta_{jL}^{E,C}& = &a_{s,j}^{E,C}c_{b,j}^{E,C},\\ 
 \gamma_{t}^{E,C}& = &(c_{s,t}^{E,C})^{*}b_{b,t}^{E,C},\\ 
 \gamma_{c}^{E,C}& = &c_{b,c}^{E,C}b_{s,c}^{E,C},
\end{array}
\qquad 
\begin{array}{rcl}
 \delta_{t}^{E,C}& = &a_{s,t}^{E,C}b_{b,t}^{E,C}, \\
 \delta_{c}^{E,C}& = &(a_{b,c}^{E,C})^{*}b_{s,c}^{E,C},\\ 
 \epsilon_{t}^{E,C}& = &(c_{s,t}^{E,C})^{*}d_{b,t}^{E,C}, \\ 
 \epsilon_{c}^{E,C}& = &c_{b,c}^{E,C}d_{s,c}^{E,C}, \\ 
 \zeta_{t}^{E,C}& = &a_{s,t}^{E,C}d_{b,t}^{E,C}, \\ 
 \zeta_{c}^{E,C}& = &(a_{b,c}^{E,C})^{*}d_{s,c}^{E,C}.
\end{array}
$$
\subsection{The Z penguin ($\alpha$=1)}
\paragraph{PC case}
\begin{eqnarray}
\nonumber \mathcal{L}_{PC}^{Z}&=& \frac{1}{(4\pi)^{2}M_{W}^{2}}\frac{g^{3}}{\cos\theta_{W}}\left\{O^{\rho}_{L}\sum_{j}\left[\alpha_{jL}+\gamma_{jL}+\zeta_{jL}+\eta_{jL}\right]\Gamma^{(i)}(x_{j})+\right. \\ \nonumber
&+&O^{\rho}_{R}\sum_{j}\left[\alpha_{jR}+\gamma_{jR}+\zeta_{jR}+\eta_{jR}\right]\Delta^{(i)}(x_{j})- \\ \nonumber
&+&\sum_{j}\left[m_{b}\left(\beta_{jR}+\epsilon_{jR}+\delta_{jR}+\theta_{jR}\right)O^{\rho}_{R}\left(G^{(i)}(x_{j})+H^{(i)}(x_{j})\right)\right.\\ \nonumber
&+& m_{s}\left(\beta_{Rj}+\epsilon_{jR}+\delta_{jR}+\theta_{jR}\right)O^{\rho}_{L}G^{(i)}(x_{j})-\\\nonumber
&-&3m_{b}\left(\beta_{jL}+\epsilon_{jL}+\delta_{jL}+\theta_{jL}\right)O^{\rho}_{R}\left(E^{(i)}(x_{j})+F^{(i)}(x_{j})\right)-\\ \nonumber
&-&\left.\left.m_{s}\left(\beta_{jL}+\epsilon_{jL}+\delta_{jL}+\theta_{jL}\right)O^{\rho}_{L}E^{(i)}(x_{j})\right]\right\}Z_{\rho}.
\end{eqnarray}
\paragraph{PB case}
\begin{eqnarray}
\nonumber \mathcal{L}_{PB}^{Z}&=& \frac{1}{(4\pi)^{2}M_{W}^{2}}\frac{g^{3}}{\cos\theta_{W}}\left\{O^{\rho}_{L}\left[\sum_{j}\alpha_{jL}\Gamma^{(i)}(x_{j})+\gamma_{t}\Gamma^{(i)}(x_{t})+\zeta_{c}\Gamma^{(i)}(x_{c})\right]+\right. \\ \nonumber
&+&O^{\rho}_{R}\left[\sum_{j}\alpha_{jR}\Delta^{(i)}(x_{j})+\gamma_{c}\Delta^{(i)}(x_{c})+\zeta_{t}\Delta^{(i)}(x_{t})\right]- \\ \nonumber
&-&3m_{b}\epsilon_{c}O^{\rho}_{R}\left[E^{(i)}(x_{c})+F^{(i)}(x_{c})\right]-m_{s}\epsilon_{c}O^{\rho}_{L}E^{(i)}(x_{c})+\\ \nonumber&+&m_{b}\epsilon_{t}O^{\rho}_{R}\left[G^{(i)}(x_{t})+H^{(i)}(x_{t})\right]+m_{s}\epsilon_{t}O^{\rho}_{L}G^{(i)}(x_{t})+\\ 
&+&m_{b}\delta_{c}O^{\rho}_{R}\left[G^{(i)}(x_{c})+H^{(i)}(x_{c})\right]+m_{s}\delta_{c}O^{\rho}_{L}G^{(i)}(x_{c})-\\\nonumber
&-&3m_{b}\delta_{t}O^{\rho}_{R}\left[E^{(i)}(x_{t})+F^{(i)}(x_{t})\right]-m_{s}\delta_{t}O^{\rho}_{L}E^{(i)}(x_{t})+ \\\nonumber
&+&\sum_{j}\left[m_{b}\beta_{Rj}O^{\rho}_{R}\left(G^{(i)}(x_{j})+H^{(i)}(x_{j})\right)+m_{s}\beta_{Rj}O^{\rho}_{L}G^{(i)}(x_{j})\right.-\\\nonumber
&-&\left.\left.3m_{b}\beta_{Lj}O^{\rho}_{R}\left(E^{(i)}(x_{j})+F^{(i)}(x_{j})\right)-m_{s}\beta_{Lj}O^{\rho}_{L}E^{(i)}(x_{j})\right]\right\}Z_{\rho}.
\end{eqnarray}
In the previous expression we have denoted with $\beta$ (which stands for a generic Greek letter) both $\beta^{E}$ and $\beta^{C}$. Obviously $\beta^{C}$ is always multiplied by a loop function whose argument is $x_{T}=M_{T}/M_{W}$. For the full structure of the loop functions and the computation in the case $\alpha\ne 1$ we refer to \cite{D'Agnolo:2009th}.
\subsection{The $\Delta F=2$ transitions}
In the following we adopt a compact notation, not specifying the arguments of the loop functions, that are uniquely determined by their coefficients and omitting the sums over the indexes of the Greek letters.
In this way we have
\paragraph{PC case}
\begin{eqnarray}
 \mathcal{L}_{PC}^{box}&=&\frac{g^{4}}{64\pi^{2}M_{W}^{6}}\left\{(\alpha_{R}+\zeta_{R}+\gamma_{R}+\eta_{R})^{2}\mathcal{O}_{RR}G_{0}+(\alpha_{L}+\zeta_{L}+\gamma_{L}+\eta_{L})^{2}\mathcal{O}_{LL}G_{0}\right.+\\\nonumber&+&(\beta_{R}+\epsilon_{R}+\delta_{R}+\theta_{R})^{2}\mathcal{O}^{'}_{R}G_{1}+\\ \nonumber
&+&(\beta_{L}+\epsilon_{L}+\delta_{L}+\theta_{L})^{2}\mathcal{O}^{'}_{L}G_{1}+(\beta_{R}+\epsilon_{R}+\delta_{R}+\theta_{R})(\beta_{L}+\epsilon_{L}+\delta_{L})\mathcal{O}^{'}_{LR}G_{1}+\\\nonumber
&+&(\alpha_{R}+\zeta_{R}+\gamma_{R}+\eta_{R})(\alpha_{L}+\zeta_{L}+\gamma_{L}+\eta_{L})\mathcal{O}_{LR}G_{0}+\\ \nonumber
&+&(\alpha_{R}+\zeta_{R}+\gamma_{R}+\eta_{R})(\beta_{R}+\epsilon_{R}+\delta_{R}+\theta_{R})\mathcal{O}^{'}_{LR}m_{b}(A+C)+\\ \nonumber
&+&(\alpha_{R}+\zeta_{R}+\gamma_{R}+\eta_{R})(\beta_{L}+\epsilon_{L}+\delta_{L}+\theta_{L})\mathcal{O}^{'}_{L}m_{b}(A+C)+ \\ \nonumber
&+&(\alpha_{L}+\zeta_{L}+\gamma_{L}+\eta_{L})(\beta_{R}+\epsilon_{R}+\delta_{R}+\theta_{R})\mathcal{O}^{'}_{R}m_{b}(A+C)+
\\ \nonumber&+&\left.(\alpha_{L}+\zeta_{L}+\gamma_{L}+\eta_{L})(\beta_{L}+\epsilon_{L}+\delta_{L}+\theta_{L})\mathcal{O}^{'}_{LR}m_{b}(A+C)\right\},
\end{eqnarray}
\paragraph{PB case}
\begin{eqnarray}
\mathcal{L}_{PB}^{box}&=&\frac{g^{4}}{64\pi^{2}M_{W}^{6}}\left\{(\alpha_{R}+\zeta_{t}+\gamma_{c})^{2}\mathcal{O}_{RR}G_{0}+(\alpha_{L}+\zeta_{c}+\gamma_{t})^{2}\mathcal{O}_{LL}G_{0}\right.+\\\nonumber&+&(\beta_{R}+\epsilon_{t}+\delta_{c})^{2}\mathcal{O}^{'}_{R}G_{1}+\\ \nonumber
&+&(\beta_{L}+\epsilon_{c}+\delta_{t})^{2}\mathcal{O}^{'}_{L}G_{1}+(\beta_{R}+\epsilon_{t}+\delta_{c})(\beta_{L}+\epsilon_{c}+\delta_{t})\mathcal{O}^{'}_{LR}G_{1}+\\\nonumber
&+&(\alpha_{R}+\zeta_{t}+\gamma_{c})(\alpha_{L}+\zeta_{c}+\gamma_{t})\mathcal{O}_{LR}G_{0}+
\\
\nonumber
&+&(\alpha_{R}+\zeta_{t}+\gamma_{c})(\beta_{R}+\epsilon_{t}+\delta_{c})\mathcal{O}^{'}_{LR}m_{b}(A+C)+\\ \nonumber
&+&(\alpha_{R}+\zeta_{t}+\gamma_{c})(\beta_{L}+\epsilon_{c}+\delta_{t})\mathcal{O}^{'}_{L}m_{b}(A+C)+ \\ \nonumber
&+&(\alpha_{L}+\zeta_{c}+\gamma_{t})(\beta_{R}+\epsilon_{t}+\delta_{c})\mathcal{O}^{'}_{R}m_{b}(A+C)+
\\ \nonumber&+&\left.(\alpha_{L}+\zeta_{c}+\gamma_{t})(\beta_{L}+\epsilon_{c}+\delta_{t})\mathcal{O}^{'}_{LR}m_{b}(A+C)\right\},
\end{eqnarray}
Here we have neglected terms of order $m_{s}$ and $p_{4}^{\mu}$, where $p_{4}^{\mu}$ is the momentum of the outgoing $s$ quark. As for the Z penguin the loop function can be found in \cite{D'Agnolo:2009th}. Note that the loop functions do not diverge.
\subsection{The photon penguin}
Here with $A_{SM}$ we have denoted the Inami-Lim function that is obtained performing the same calculation in the gaugeless limit of the Standard Model \cite{Inami:1980fz}. The convention for the Greek letters and the superscripts $E$ and $C$ is the same adopted for the Z penguin.
\paragraph{PC case}
\begin{eqnarray}
 \mathcal{L}^{\gamma}_{PC}&=&\frac{e}{(4\pi)^{2}}\frac{g^{2}}{2M_{W}^{2}}\sum_{j}A_{SM}(x_{j})\left\{\left(\beta_{jR}+\epsilon_{jR}+\delta_{jR}+\theta_{jR}\right)O_{1}+\right. \\ \nonumber
&+& \left.\left(\beta_{jL}+\epsilon_{jL}+\delta_{jL}+\theta_{jL}\right)O_{2}\right\}.
\end{eqnarray}
\paragraph{PB case}
\begin{eqnarray}
\mathcal{L}^{\gamma}_{PB}&=&\frac{e}{(4\pi)^{2}}\frac{g^{2}}{2M_{W}^{2}}\left\{\sum_{j}\left[\beta_{jR}O_{1}+\beta_{jL}O_{2}\right]A_{SM}(x_{j})+\right. \\ \nonumber
&+& \left.\epsilon_{c}O_{1}A_{SM}(x_{c})+\epsilon_{t}O_{2}A_{SM}(x_{t})+\delta_{c}O_{2}A_{SM}(x_{c})+\delta_{t}O_{1}A_{SM}(x_{t})\right\}.
\end{eqnarray}

\end{document}